# An Alternative Route to D-wave Superconductivity: the Charge Channel

D. Z. Liu and K. Levin

*James Franck Institute, University of Chicago, Chicago, IL 60637*

(Submitted to Physical Review Letters on 12 August 1995)

We demonstrate that a $d_{x^2-y^2}$ superconducting pairing is naturally associated with the screened Coulomb interaction $V_{eff}(\mathbf{q},\omega)$ within the $CuO_2$ plane of high $T_c$ superconductors. This pairing instability arises (via the dielectric constant) from wave-vector structure in $V_{eff}(\mathbf{q},\omega)$ at $\mathbf{q}=(\pi,\pi)$, which is associated with 2D Van Hove and local field effects. Our results, which are independent of the bandstructure, suggest that direct Coulomb intereactions are significant (when compared to the spin channel) and will, at the least, enhance any other underlying mechanism for $d_{x^2-y^2}$ wave pairing.

There is a growing body of evidence to support the claim that, at least in some of the high $T_c$ cuprates, the superconducting order parameter has $d$-wave symmetry [1]. This evidence, although by no means conclusive, is strongest for one particular compound YBCO. While, it has been argued theoretically [1] that this d-wave state may be driven through exchange of antiferromagnetic spin fluctuations, neutron experiments above $T_c$ in YBCO repeatedly fail [2] to observe all but very weak antiferromagnetic peaks associated with the optimally doped material.

The absence of significant spin fluctuation effects has motivated us to investigate the charge channel as a possible source for $d$-wave attraction. Other workers have considered charge fluctuation induced superconductivity primarily via multi-band or excitonic effects, though, within this channel, none of these were found to lead to $d$- wave pairing [3]. In this paper we investigate an alternative, albeit one band scenario, based on the charge dynamics and associated wave vector structure. We demonstrate that the screened Coulomb interaction

$$V_{eff}(\mathbf{q},\omega) = V_o(\mathbf{q})/\epsilon(\mathbf{q},\omega) \qquad (1)$$

while repulsive in $\mathbf{q}$ space, is attractive in the vicinity of nearest neighbor spatial separations in real space. Here $V_o = 2\pi e^2/\epsilon_s q$ for a 2D system [4] (and $\epsilon_s$ is the static dielectric constant of the material). Moreover, this structure leads naturally to a $d_{x^2-y^2}$-wave order parameter symmetry [5].

It was pointed out by Kohn and Luttinger [6] more than three decades ago that superconductivity may arise from repulsive Coulomb effects. This attraction is associated with the Kohn anomaly or singularity in the dielectric constant at momentum transfer $q = 2k_F$ which gives rise to a long range oscillatory interactions (Friedel oscillations) in co-ordinate space. Because these studies were performed on a three dimensional electron gas, the resulting superconducting instability was weak and in a high angular momentum channel.

In order to generalize the Kohn Luttinger picture in the presence of a lattice, we follow earlier work by Sham [8] and Pandey et al [9] who presented calculations of the dielectric constant which included local field corrections. The screened Coulomb potential defined in Eq. (1) can be written as

$$V_{eff}(\mathbf{q}+\mathbf{G}) = V_o(\mathbf{q}+\mathbf{G}) \qquad (2)$$
$$\times [1+\sum_{\mathbf{G}'} \chi(\mathbf{q}+\mathbf{G},\mathbf{q}+\mathbf{G}')V_o(\mathbf{q}+\mathbf{G}')]$$

Here $\mathbf{G}$ is a reciprocal lattice vector and $\mathbf{q}$ is in the first Brillouin zone. For notational simplicity we drop the $\omega$ indices. If we define $\chi_{\mathbf{G}\mathbf{G}'}(\mathbf{q}) = \chi(\mathbf{q}+\mathbf{G},\mathbf{q}+\mathbf{G}')$ then we have

$$\chi_{\mathbf{G}\mathbf{G}'} = \chi^o_{\mathbf{G}\mathbf{G}'} + \sum_{\mathbf{G}''} \chi^o_{\mathbf{G}\mathbf{G}''} V_o(\mathbf{q}+\mathbf{G}'')\chi_{\mathbf{G}''\mathbf{G}'} \qquad (3)$$

within the Random-Phase-Approximation (RPA). The noninteracting polarizability is given by

$$\chi^o_{\mathbf{G}\mathbf{G}'}(\mathbf{q},\omega) = 2 \sum_{\mathbf{k}} \frac{n_{\mathbf{k}} - n_{\mathbf{k+q}}}{\omega + E_{\mathbf{k}} - E_{\mathbf{k+q}} + i\delta} \qquad (4)$$
$$\times <\mathbf{k}|e^{-i(\mathbf{q}+\mathbf{G})\cdot \mathbf{r}}|\mathbf{k+q}>$$
$$\times <\mathbf{k+q}|e^{i(\mathbf{q}+\mathbf{G}')\cdot \mathbf{r}}|\mathbf{k}>$$

where the basis wavefunctions

$$\psi_{\mathbf{k}}(\mathbf{r}) = <\mathbf{r}|\mathbf{k}> = \frac{1}{\sqrt{N}} \sum_l C_{\mathbf{k}} e^{i\mathbf{k}\cdot \mathbf{R}_l} \phi(\mathbf{r} - \mathbf{R}_l) \qquad (5)$$

are the Bloch wavefunctions and $\phi(\mathbf{r}-\mathbf{R}_l)$ are the associated Wannier wavefunctions at lattice site $\mathbf{R}_l$. Here $C_{\mathbf{k}}$ is a renormalization constant. After some algebra, Eq.(4) can be written as:

$$\chi^o_{\mathbf{G}\mathbf{G}'}(\mathbf{q},\omega) = 2 \int_{BZ} \frac{d^2\mathbf{k}}{(2\pi)^2} \frac{n_{\mathbf{k}} - n_{\mathbf{k+q}}}{\omega + E_{\mathbf{k}} - E_{\mathbf{k+q}} + i\delta} \qquad (6)$$
$$\times |C_{\mathbf{k}}|^2 |C_{\mathbf{k+q}}|^2$$
$$\times \sum_{l_1 l_2} e^{i\mathbf{k}\cdot(\mathbf{R}_{l_1}-\mathbf{R}_{l_2})} t_{l_1}(\mathbf{q}+\mathbf{G}) t^*_{l_2}(\mathbf{q}+\mathbf{G}')$$

where

$$t_l(\mathbf{k}) = \int d^2\mathbf{r}\,\phi(\mathbf{r})\phi^*(\mathbf{r}+\mathbf{R}_l)e^{-i\mathbf{k}\cdot\mathbf{r}} \qquad (7)$$

We apply this approach to the cuprates by incorporating standard parameterizations of the band structure in a tight binding square lattice:

$$E_\mathbf{k} = -2t[\cos(k_x)+\cos(k_y)] + 4t'\cos(k_x)\cos(k_y) \qquad (8)$$

where $t$ is the nearest and $t'$ the next-nearest-neighbor hopping energy. The Wannier wavefunctions can be fitted to a Gaussian, whose width is adjusted to correspond to the known [7] values of the ratio $t'/t$. One can then determine the non-interacting polarizability given in Eq.(6), and from this the dielectric constant and effective interaction are deduced following Eqs. (2) and (3).

To place our results in the proper perspective, in Figure 1 we plot the wave-vector dependence of the calculated, static Coulomb interaction $V_{eff}(\mathbf{q})$ for the standard YBCO bandstructure as compared with the dynamical spin susceptibility; the latter is derived from fits to low $\omega$ neutron experiments [2]. By normalizing both curves to the same average value, this figure represents a plot of the relative importance of wave-vector structure in the spin (dotted line) and charge (solid line) channels [10]. Under the assumption that the electrons couple with equal strength to each channel, and with the same "Debye" frequency, a weak coupling calculation would lead to a considerably higher ($d_{x^2-y^2}$-wave) $T_c$ for the charge induced interactions.

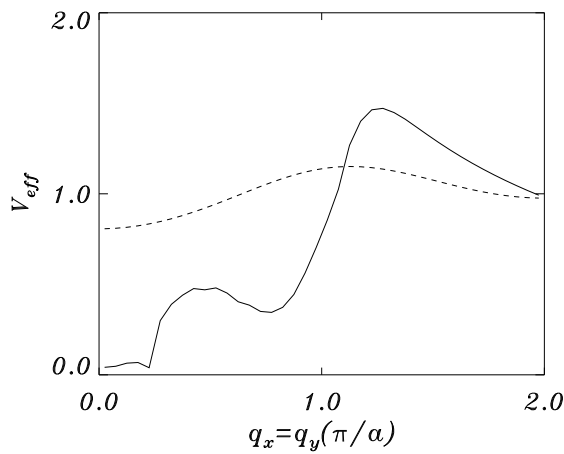

FIG. 1. Comparison of the calculated interaction in the charge channel (screened Coulomb pairing potential) in YBCO (at $E_F = -1.6t$) (solid line) with that deduced phenomenologically in the spin-channel (dashed line).

To understand the important role of the lattice, we first compare with the 2D electron gas RPA case, at $\omega = 0$, where

$$\epsilon(\mathbf{q},\omega) = 1 - V_o(\mathbf{q})\chi^o(\mathbf{q},\omega) \qquad (9)$$

and

$$\chi^o(\mathbf{q},\omega) = -2\int\frac{d^2\mathbf{k}}{(2\pi)^2}\frac{n_{\mathbf{k+q}}-n_\mathbf{k}}{\omega - E_{\mathbf{k+q}}+E_\mathbf{k}+i\delta}, \qquad (10)$$

where $E_\mathbf{q} = \hbar^2 q^2/2m$. In the static limit this latter function exhibits a Kohn anomaly as seen in the dashed line in the inset of Fig. 2. The resulting screened Coulomb interaction shown in the main portion of the figure (dashed line) starts to rise at $q=2k_F$, forming a peak at $q>2k_F$. This dashed line was found as a limit of our calculational scheme in the case of extremely low electron filling (where the electronic dispersion is quadratic). For more general values of the electron count, and for the usual bandstructure associated with LSCO, the screened interaction, derived from Eqs.(2) and (3) is given by the solid line in the main portion of the figure. The solid line in the inset of Fig. 2 represents the deduced form for the associated $\chi^o(\mathbf{q})$ which essentially includes local field corrections, but fitted to the form of Eq.(9). This susceptibility should be contrasted with the Lindhard function which takes the form of Eq.(10) but in which the integration is limited to the first Brillouin zone, so that local field terms are absent. It should be noted by comparing Fig. 1 and Fig. 2 that there only small differences between the behavior of the screened Coulomb interaction in the two different cuprates.

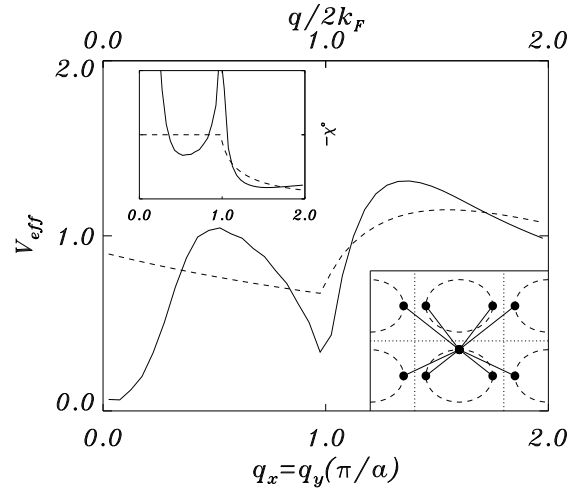

FIG. 2. Comparison of the effective Coulomb potential in LSCO at the Van-Hove point (solid line) with the case of a 2D free electron gas (dashed line). The upper inset plots the effective susceptibility for these two cases, illustrating the effect of the Van-Hove singularity on the Kohn anomaly. The lower inset demonstrates how Van-Hove effects are enhanced by Umklapp processes.

The principal difference between the full lattice calculation (solid line) and the electron gas limit (dashed line) can be seen from Fig. 2 to be associated with both the small $\mathbf{q}$ and $\mathbf{q}=(\pi,\pi)$ regimes. The latter is the more important, since the former is associated with very long distance Friedel oscillations, which are not expected to play a role in the superconductivity. The origin of the structure around $(\pi,\pi)$ in both Fig. 1 and Fig. 2 is a "two particle" Van-Hove effect which may be viewed as a "depletion" in the intensity of the screened interaction (relative to the dashed line) as a consequence of a maximum in the diectric function. This in turn arises from the fact that in a 2D lattice, the density of states in momentum space has divergences at the Van-Hove points of the Brillouin zone edge. For a square lattice (appropriate to the cuprates), these points are at $(\pm\pi, 0)$ and $(0, \pm\pi)$. In this way when the momentum differences between the initial and final states in an electric field induced polarization process correspond to the separation of two Van-Hove points, one expects a peak in the dielectric constant. Since most Fermi surfaces do not necessarily intersect the Van-Hove points, the peak position is determined by the momentum transfer between two points on the Fermi surface which are in the neighborhood of the Van-Hove points. Here we see that local field corrections which enter via Umklapp processes are important so that the "two particle" Van Hove effect necessarily includes contributions from momentum space beyond the first Brillouin zone. These processes, which enhance the Van Hove contributions further, are shown in the lower inset of Fig. 2 [11].

The above results lead to important implications for the Kohn Luttinger model for superconductivity, when generalized to a 2D lattice. It should be stressed that the original calculations were based on a weak coupling model, as well as the assumption that Migdal's theorem is valid for electronic mechanisms. The latter is a highly complex issue which we will not adress here [12] since the issue of primary interest is the order parameter symmetry rather than magnitude of $T_c$. For these approximations it is relatively straightforward to deduce the symmetry of the order parameter.In the case of an electron gas, the screened Coulomb interaction depends only on $|\mathbf{q}|$. Because there is no directionality, there can be no $d_{x^2-y^2}$-wave instability. By contrast, in the 2D lattice, the screened interaction depends on the wave-vector direction. Moreover, there is a maximum slightly above $(\pi,\pi)$ as shown in the first two figures. In this way we consistently find that the resulting superconducting order parameter has $d_{x^2-y^2}$-wave symmetry independent of the electronic filling and the band structure . A typical gap function is shown by the solid line in the inset of Fig. 3.

An issue of primary interest in understanding the pairing mechanism is determining an associated dynamical energy scale, which can lead to sufficiently high $T_c$. This is all the more problematic in the context of the low values of the boson coupling constants which appear in transport experiments [13]. For the spin fluctuation case, the energy scale observed from neutron data [2] is comparable to that of phonons, and thus it may not be sufficiently high.

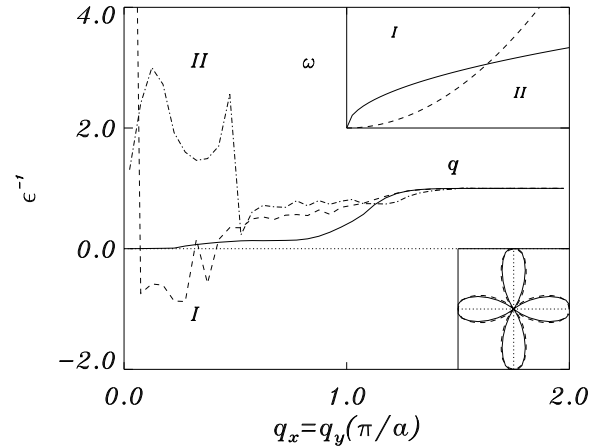

FIG. 3. Inverse of dielectric constant for YBCO (at $E_F = -1.6t$) as a function of frequency: solid line $\omega = 0$; dashed line, $\omega = 3.2t$; dash-dotted line, $\omega = 6.4t$. The corresponding superconducting gaps of the solid and dashed cases are shown in the lower inset. Upper inset plots the dispersion relation for the plasmon mode (solid line) and the edge of the particle-hole continuum (dashed line).

At finite frequency, there are two types of relevant excitations contained in the screened Coulomb interaction. One is the particle-hole "exciton" or continuum and the other, the usual plasmon collective mode ($\omega_p$). The dielectric constant vanishes at the latter while its inverse ($\epsilon^{-1}$) passes through zero at the continuum edge. ($\omega_{ex}$). As shown in the upper inset of Fig. 3, there are two regimes in the parameter space called I,II. At fixed q, the first satisfies $\omega_p > \omega_{ex}$ and the second $\omega_p < \omega_{ex}$. The dashed line of Fig. 3 shows the behavior of $\epsilon^{-1}$ for moderate, real frequencies ($\omega = 3.2t$) which correspond to region I. While there is a sign change at small wave-vectors, the feature at $(\pi,\pi)$ still dominates the interaction and the resulting (finite frequency) order parameter is essentially the same as for the static limit, as can be seen from the lower inset. At higher frequency ($\omega = 6.4t$), corresponding to II, the behavior is plotted as the dot-dashed line of Fig. 3. For these high frequencies, the $d$ wave state is no longer stable. In this way one deduces a characteristic pairing energy for $d$ wave superconductivity, which corresponds to that frequency where the plasmon mode enters the particle-hole continuum. For YBCO, we estimate this cutoff frequency $\omega_{cutoff} \approx 4t$, which for reasonable values of $t$ ranges from 250 to 1000 meV. Clearly a full dynamical Eliashberg calculation should ultimately be performed to incorporate this finite frequency effect, although this is also subject to earlier concerns about Migdal's theorem. Nevertheless the present calculations demonstrate that the energy scale for the charge channel should be considered to be of the order of hundreds of meV, as contrasted with tens of meV in the spin channel. Furthermore, this behavior appears consistent with Uemura's scaling plots [14].

Finally, we note that the presence of **q** structure in $\epsilon(\mathbf{q},\omega)$ also has important implications for the screened electron-phonon interaction and its contribution to any superconducting pairing. Assuming that the phonon propagator is relatively isotropic so that it can be approximated by a coupling constant of strength $\lambda_p$, the total effective interaction is then

$$V_{eff} = V_o/\epsilon - \lambda_p/\epsilon \qquad (11)$$

While, in principle, pseudopotential effects should be included, we may incorporate these phenomenologically through an effectively stronger $\lambda_p$. In Fig. 4, we present the calculated superconducting gap for various values of the phonon coupling strength. The figure illustrates the extreme case where the Fermi energy is close to the Van Hove singularity, so as to emphacize Van Hove effects. In the absence of phonons, as well as for the case of small $\lambda_p = 0.5$, a similar $d_{x^2-y^2}$-wave gap is obtained. With increasing coupling the gap first evolves to a twelve node d-wave state ($\lambda_p = 1$), then to a highly anisotropic s-wave state ($\lambda_p = 2$) and ultimately ($\lambda_p = 5$) to a somewhat more isotropic s-wave symmetry. The Van Hove singularity plays an important role in, whenever possible, maximizing the gap values at the four Van Hove points. Somewhat away from the Van Hove singularity, we find that an intermediate $d_{xy}$ state arises before the transition to s-wave symmetry. The behavior discussed here should be contrasted with the complementary calculations in Ref. [15], where the role of Van Hove effects on the electron phonon interaction were also studied [16].

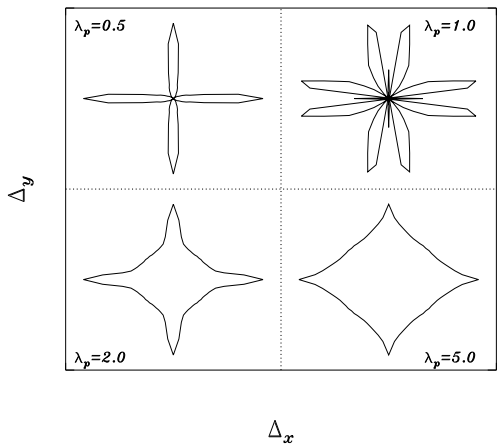

FIG. 4. Superconducting gap for YBCO at the van-Hove point for various values of the electron- phonon coupling strength ($\lambda_p$).

In summary, we find important wave vector structure in the dielectric constant and therefore in the screened Coulomb interaction at $\mathbf{q} = (\pi,\pi)$. This structure is associated with the 2D Van Hove singularities in conjunction with local field corrections, and leads to $d_{x^2-y^2}$ wave pairing. This physical picture should be viewed as a charge analogue of the standard scenario used to deduce d-wave superconductivity from the spin channel. It is the amplitude and position of **q** structure in the interaction which determines the nature and degree of attraction in real space. Future calculations will extend this work to the multi-band case. Whether this provides all or only a portion of the superconducting attraction, our calculations demonstrate that the direct Coulomb channel will always act to enhance any (other) underlying propensity for d- wave superconductivity.

We thank M. Norman, J. Maly and X.C. Xie for helpful discussions. This work is supported by the National Science Foundation (DMR 91-20000) through the Science and Technology Center for Superconductivity.